\documentclass[10pt,twocolumn,letterpaper]{article}

\usepackage[pagenumbers]{cvpr} 

\usepackage{graphicx}
\usepackage{amsmath,bm}
\usepackage{amssymb}
\usepackage{booktabs}

\usepackage[pagebackref,breaklinks,colorlinks]{hyperref}

\usepackage[capitalize]{cleveref}
\crefname{section}{Sec.}{Secs.}
\Crefname{section}{Section}{Sections}
\Crefname{table}{Table}{Tables}
\crefname{table}{Tab.}{Tabs.}

\def\cvprPaperID{*****} 
\def\confName{CVPR}
\def\confYear{2022}

\begin{document}

\title{Multi-organ Segmentation Network with Adversarial Performance Validator}

\author{Haoyu Fang$^{*}$\\
New York University\\
New York, NY $11201$, USA\\
{\tt\small haoyu.fang@nyu.edu}
\and
Yi Fang\\
New York University Abu Dhabi\\
Abu Dhabi, PO $129188$, UAE\\
{\tt\small yfang@nyu.edu}
\and
Xiaofeng Yang\\
Winship Cancer Institute of Emory University\\
Atlanta, GA $30322$, USA\\
}
\maketitle
\begin{abstract}
   CT organ segmentation on computed tomography (CT) images becomes a significant brick for modern medical image analysis, supporting clinic workflows in multiple domains. Previous segmentation methods include 2D convolution neural networks (CNN) based approaches, fed by CT image slices that lack the structural knowledge in axial view, and 3D CNN-based methods with the expensive computation cost in multi-organ segmentation applications. This paper introduces an adversarial performance validation network into a 2D-to-3D segmentation framework. The classifier and performance validator competition contribute to accurate segmentation results via back-propagation. The proposed network organically converts the 2D-coarse result to 3D high-quality segmentation masks in a coarse-to-fine manner, allowing joint optimization to improve segmentation accuracy. Besides, the structural information of one specific organ is depicted by a statistics-meaningful prior bounding box, which is transformed into a global feature leveraging the learning process in 3D fine segmentation. The experiments on the NIH pancreas segmentation dataset demonstrate the proposed network achieves state-of-the-art accuracy on small organ segmentation and outperforms the previous best. High accuracy is also reported on multi-organ segmentation in a dataset collected by ourselves.
\end{abstract}

\section{Introduction}

As one of the preferred radiotherapy for H\&N cancer, intensity-modulated radiation therapy (IMRT), employing advance shape radiation dose for the complex H\&N anatomy and pathology, significantly reduces the parotid dose that has severe long-term side effect. However, the effectiveness of IMRT is strongly associated with accuracy of exposure of organs-at-risk (OARs) to avoid unnecessarily high dose. Conventionally, the OARs segmentation is performed by professional oncologists and dosimetrists and cost great manual effort as well as time. 

To alleviate aforementioned issue, auto-segmentation for OARs in H\&N areas based on CT scan images is proposed, which should provide precise segmentation against many challenges (i.e. inter-patient variability and relatively low CT soft tissue contrast and morphological complexity resulted from a large number of anatomical structures in small area \cite{tong2018fully}). With the progress of advance convolution neural networks \cite{imagenet,u-net} applying in computer vision tasks (e.g. object classification and recognition, object detection and image segmentation etc.), increasing efforts to introduce deep learning networks on organ segmentation in medical data are made, which results in ascending of segmentation accuracy \cite{JHU,u-net}. 

The current deep-learning-based organ segmentation methods roughly consist of 2D segmentation networks \cite{JHU,u-net} and 3D segmentation networks \cite{3DRPN,JHU2,3du-net}. 2D segmentation methods process mainly on 2D slices, which do not contain information in Z-axis, resulting in geometric information loss. 3D segmentation generates a 3D mask from volume information and many approaches based on it achieve state-of-the-art performance. However, due to a large number of latent parameters in 3D networks, 3D segmentation networks are expensive in computation cost. When going to multi-organ segmentation tasks, approaches based on 3D networks can consume computation resource more than hardware (e.g. GPU) support. Besides, the low CT soft tissue contrast challenges most existing automated segmentation methods by confusing the network to distinguish boundaries of specific OARs and their surrounding tissues. 
To alleviate these challenges, we proposed a 2D-3D hybrid segmentation network in coarse-to-fine manner with a novel Adversaria Performance Validator, which refines segmentation by penalizing residual organ information that remains after a soft mask-out process, to achieve accurate and detailed segmentation for the OARs. We also leverage location prior knowledge of organs as a global spatial constraint to generate 3D high-quality segmentation mask.

\begin{figure}
    \centering
    \includegraphics[scale=0.4]{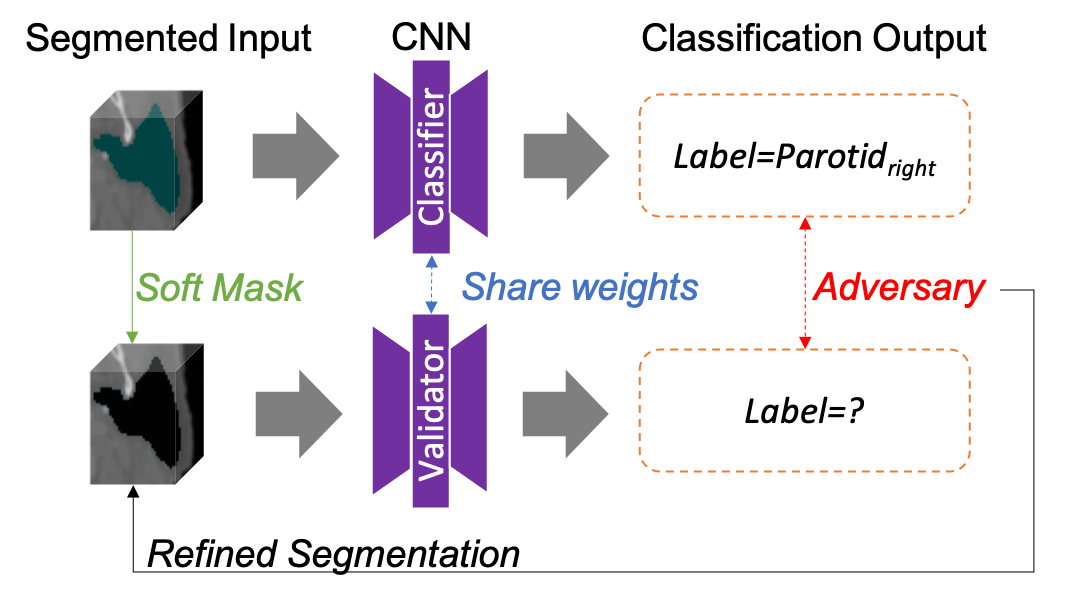}
    \caption{Illustration of Adversarial Performance Validator (APV). The classifier is trained to provide an accurate predicted label whereas it is trained for the validator impossible to predict right organ class when it is fed by a masked volume that contains little information of the target organ after a masking operation based on precise segmentation results.  Details will be introduced in  Section 3.4.}
    \label{fig1-1}
\end{figure}

\noindent\textbf{Related works}

In current clinical radiotherapy practice, CT is commonly acquired for treatment planning. Organ segmentation on CT scans by manual efforts was the main method in the early modern medical image analysis. Although organ segmentation by professional radiation oncologist provided high segmentation accuracy, the method is too labor-intensive, time-costly and impractical to support most clinical workflows, motivating an urgent need for auto/semi-auto organ segmentation approaches. Segmentation methods based on hand-craft organ-specific image features (e.g. threshold \cite{threshold}, growing region \cite{region_grow1,region_grow2} and weighted graph \cite{graph_cut1, graph_cut2}) achieved relatively good performance. However, with the rapid development of modern computer vision methods which typically utilize the representational power of deep learning, especially convolutional neural networks \cite{segnet, u-net, vnet, FCN}. Many conventional methods in organ segmentation based on handcrafted image feature \cite{threshold,region_grow1,region_grow2,graph_cut1} have been replaced by deep neural networks which typically yields higher segmentation accuracy. 

Deep neural networks for single organ segmentation on CT scans are fed with 3D volumetric data, triggering two types of solutions including 2D networks which process on 2D CT slices  \cite{JHU,2dorgan1,2dorgan2}, and 3D networks which take CT volumes as input directly \cite{3du-net,3dorgan1,vnet}. As one of the state-of-the-art approaches following 2D solution, \cite{JHU} proposed a coarse-to-fine deep learning framework with the recurrent saliency transformation module taking CT slices from three orthogonal planes as input. The proposed recurrent saliency transformation network repeatedly transforms the segmentation probability map from previous iterations as spatial priors in the current iteration, which organically relate coarse segmentation on a large scale and fine segmentation on a small scale. Among 3D organ segmentation approach, \cite{3du-net} as an early-stage work extended the previous 2D U-net framework \cite{u-net} by introducing 3D operations to the whole networks. Based on the architecture of the aforementioned 3D U-net, \cite{vnet} proposed residual structures, which established short-term skipping connection among all stages of the framework, resulting in an improvement of single organ segmentation.

Multi-organ segmentation demonstrates more practicability in various clinical applications comparing the single organ segmentation. Common multi-organ segmentation methods roughly consist of registration-based and registration-free approaches. Some of the registration-based methods involve statistical models \cite{statistic_model1,statistic_model2} obtained by co-registering images and their anatomical correspondences in training data, which construct the shape distribution \cite{statistic_model3} or anatomy appearances \cite{statistic_model4} and will be fitted into new images to generate segmentation masks. Other approaches apply Multi-atlas Label Fusion (MALF) module \cite{MALF1,MALF2,MALF3,MALF4} to register template images in training data to new images and generate new segmentation masks via combining propagated reference segmentation results. However, both statistic-based and MALF-based methods are restricted by the challenges in medical images registration, caused by strong diversity of human organs among patients (i.e difference of size, shape, appearance etc.) due to natural variability, disease status, and other factors. 

The challenges in medical image registration trigger the development in advance registration-free approaches. Some of the methods introduce hand-crafted organ-specific features (e.g. thresholds, density distribution, weighted graph and etc.) to segment organs, but the recent approaches have applied classifiers based on machine learning (ML) (typically Deep Neural Networks (DNN)) frameworks to flexibly and robustly selecting image features \cite{DL1,DL2}, achieving the state-of-the-art segmentation accuracy. Introducing majority-voting label fusion, \cite{deepmulti1} segmented 19 organs by feeding 2D slices in axial, sagittal and coronal views to deep convolutional neural networks and combining segmentation results from the three views. \cite{deepmulti2} segmented seven organs via a hierarchical network based on 3D U-net \cite{3du-net}. \cite{deepmulti3} proposed a level-set-based segmentation network fed with organ probability maps as features, which are generated by a 3D Fully Convolution Networks (FCN). However, applications of DNN as an entire or partial pipeline of a multi-organ segmentation architecture pose more significant challenges due to the need of processing large volumetric CT scans under limited computational memory resource than the aforementioned single segmentation. One strategy is to input smaller image patches into the networks \cite{smallimage1, vnet, smallimage2} and the other is to constrain depth of the deep networks \cite{shallownn}. Efforts to handle with the memory limitation have been laid, but still remaining the constrain challenging. An additional challenge for multi-organ segmentation is that weighting of the losses for various organs with high imbalance may result in failures of multi-organ segmentation.

\section{Methods}

\subsection{Prior bounding box}
In spite of the variety of human organs in shapes and locations among individual patients, human organs from different people still demonstrate spatial properties in common: a specific organ appears at a roughly fixed position and has scales varying in a predictable range. Therefore, the statistical observation of target organ's position and shape in CT scans gives a geometry clue for CNN network to learn integrated feature representation for the organ.

Before training the network, we conduct statistic of organ position and shape from the ground truth in the training examples. For an arbitrary organ belonging to class $n$, we measure the upper bound and lower bound of target organs along with $X, Y, Z$ axis for each samples and select the maximum and minimum boundaries over the entire training data as the prior bounding box $B_n \in \bm{B} = \{B_n|n=1,2,3..,N\}$:  
\begin{equation}
   B_n = (x_{n,cen},y_{n,cen},z_{n,cen},w_n,l_n,h_n),
    \label{eq2.1_1}
\end{equation}
where $n$ is the organ class index, $x_{n,cen},y_{n,cen},z_{n,cen}$ are the coEmoryordinates of the center voxel of bounding box, and $w_n,l_n,h_n$ are width, length and height of the bounding box respectively. Note that each prior bounding box is supposed to contain one specific organ within the boundary box. In addition, a prior feature representation of organ position and shape in CT volume is learned by feeding values of the prior bounding box into a neural network $f[\cdot | \eta]$, which will be utilized by 3D fine-scale segmentation network (described in Section 3.2) as a global feature. 
\begin{equation}
   S_n =f[ B_n | \eta],
    \label{eq2.1_2}
\end{equation}
where $\eta$ is the network parameter, and $S_n$ is the prior feature of organ belonging to the $n-th$ class.

\subsection{3D Network}

\subsection{Coarse-to-Fine Network}
\begin{figure*}[htb]
    \centering
    \includegraphics[scale=0.5]{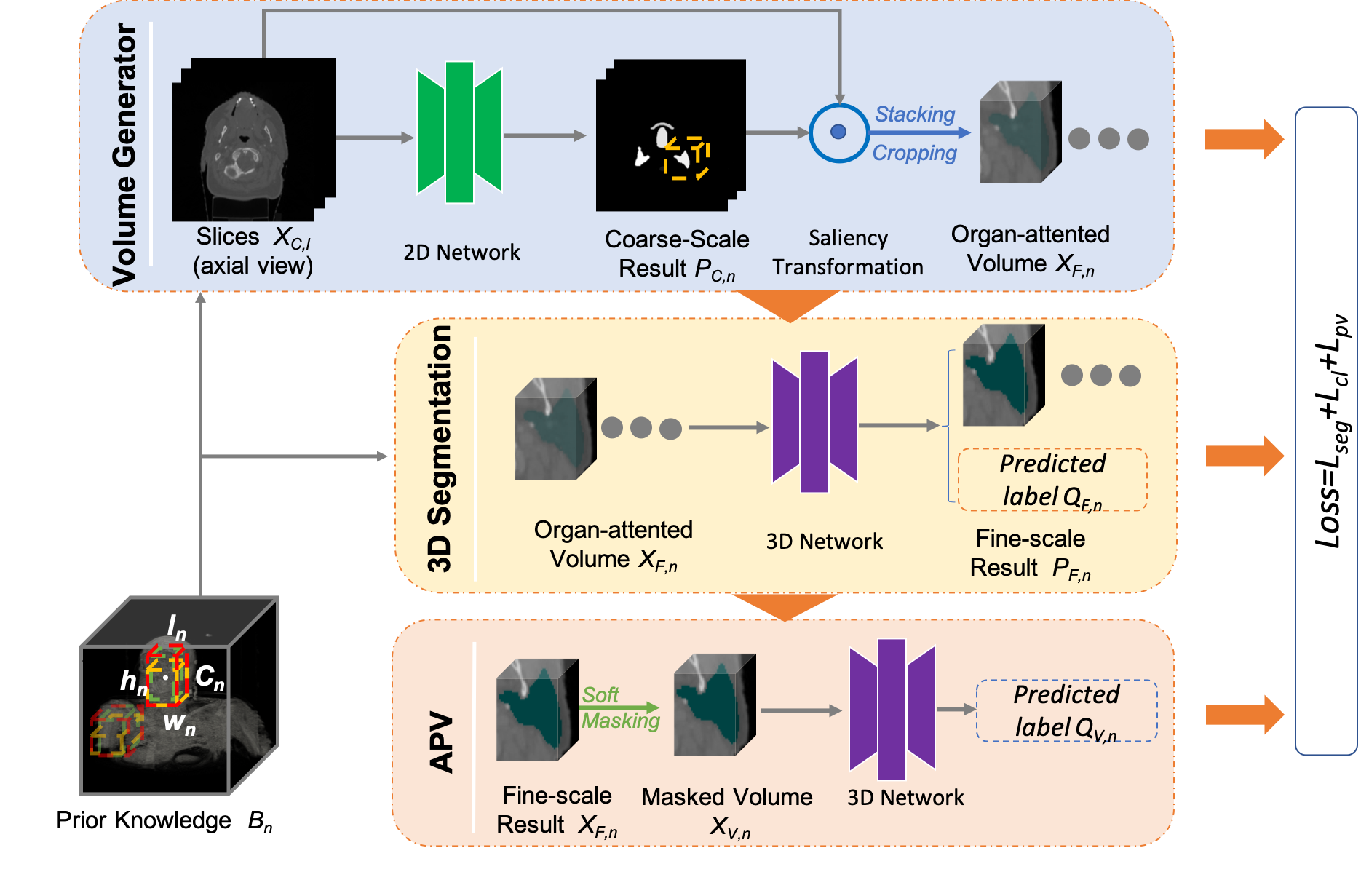}
    \caption{Pipeline of the organ segmentation framework. The blue box represents the framework of the sub-volume generator, the yellow box represents the 3D segmentation network and the Adversarial Performance Validator (APV) is illustrated by the red box. Prior knowledge assists in cropping operation within sub-volume generator and feature learning in both 3D segmentation network and the APV module.}
    \label{fig2-1}
\end{figure*}

Fig. \ref{fig2-1} illustrates the pipeline of the proposed networks, consisting of a prior bounding box capturing operation, a 3D sub-volume generator, a 3D fine-scale segmentation network and an adversarial performance validator. The input of the network is a CT volume denoted by $X$ with size $W \times H \times L$ and its label correspondence $Y$, which contains $N$ target organs' annotations. The output are supposed to be a volume $Z$ with the same dimension as $X$. Denote $Y$ and $Z$ as the set of foreground voxels in the ground-truth and prediction, i.e., $Y = \{i | y_i = n\}$ and $Z = \{i | z_i = n\}$, where $n$ represents organ category index. 

\subsection{Adversarial Performance Validator (APV)}
Due to the similarity of texture feature representations between target organ and background tissue in medical images, sometimes the segmentation network can not generate perfect segmentation mask for the specific organ target. An Adversarial Performance Validation (APV) network is introduced after the 3D segmentation network to refine segmentation results by providing more local details, especially in boundary areas (shown in Fig. \ref{fig3-1} and Fig. \ref{fig3-2}).

Fig. \ref{fig1-1} and Fig. \ref{fig2-1} demonstrate that our pipeline consists of two weights-sharing classification architectures: one is implemented within the 3D segmentation network (named the organ classifier) and the other is in the proposed APV. The organ classifier aims to find out regions that help to recognize classes and the APV is supposed to guarantee that all regions contributing to the classification decision will be included in the network’s segmentation \cite{gain}. In another perspective, the organ classifier provides an accurate predicted label while a masking operation based on precise segmentation results should make the APV impossible to predict right organ class when fed by a masked volume that contains little information of the target organ. 

We then use the probability map $P_{F,n}$ after the fine-scale segmentation to generate a soft mask applying on the original input and obtain a masked volume $X_{V,n}$ using Eq. \ref{eq3.4_1} 
\begin{equation}
    X_{V,n} = P_{F,n}-(T(P_{F,n})\odot X_{F,n} ),
    \label{eq3.4_1}
\end{equation}
where $M(\cdot)$ is a soft masking function with thresholding. Applying Sigmoid function in $T(\cdot)$ makes Eq. \ref{eq3.4_1} derivable.
\begin{equation}
    M(P_{F,n}) = \frac{1}{1+exp(-w(P_{F,n}-\sigma))},
    \label{eq3.4_2}
\end{equation}
where $sigma$ is a threshold matrix in which all elements equal to $sigma$. $w$ is the scale parameters ensuring $M(P_{F,n})_{i,j,k}$ approximate 1 when value of voxel $(i,j,k)$ in $P_{F,N}$ is larger than $\sigma$ or 0 otherwise.

$X_{V,n}$ is then fed to performance validator. To enforce the APV remain low classification scores for the target organ classes, the masked volume should convey ideally no information referring to the target class, which reflects an accurate segmentation. Therefore, a loss function named Performance Validation Loss is designed as Eq. \ref{eq3.4_3} to minimize the prediction score for target organ class:
\begin{equation}
    Loss_{apv} = min(max(s^n(X_{V,n}))),
    \label{eq3.4_3}
\end{equation}
where $s^c(\cdot)$ denotes the prediction score that $X_{V,n}$ belongs to the $n-th$ class. Meanwhile, the total loss of the proposed networks is thus defined as a weighted summation of segmentation loss, classification loss and performance validation loss in Eq. \ref{eq3.4_4}:
\begin{equation}
    Loss_{total} = \alpha Loss_{seg} + \beta Loss_{cl} + \gamma Loss_{pv},
    \label{eq3.4_4}
\end{equation}
where $L_{seg}$ is the summation of coarse-scale and fine-scale segmentation loss, and $L_{cl}$ is loss for multi-class classification (cross entropy loss is used in this paper). The $\alpha, \beta$ and $\gamma$ are weighting parameters.  With the competition between the organ classifier and the APV, the segmentation network is automatically manipulated to generate accurate segmentation masks simultaneously to ensure that the masked volume contains no information referring to target organ class. 
\begin{figure*}
    \centering
    \includegraphics[scale=0.45]{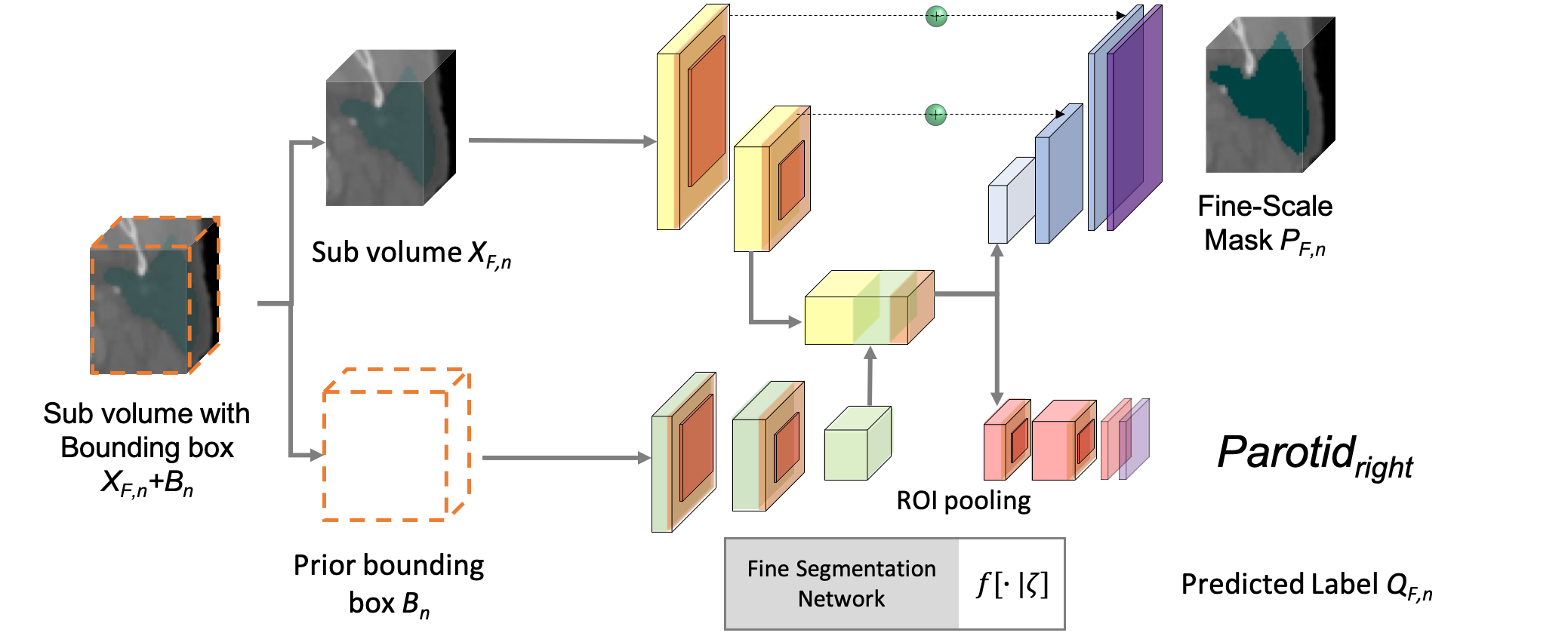}
    \caption{The 3D segmentation network.}
    \label{fig2-3}
\end{figure*}

\subsection{3D Sub-volume Generator}

\begin{figure*}
    \centering
    \includegraphics[scale=0.4]{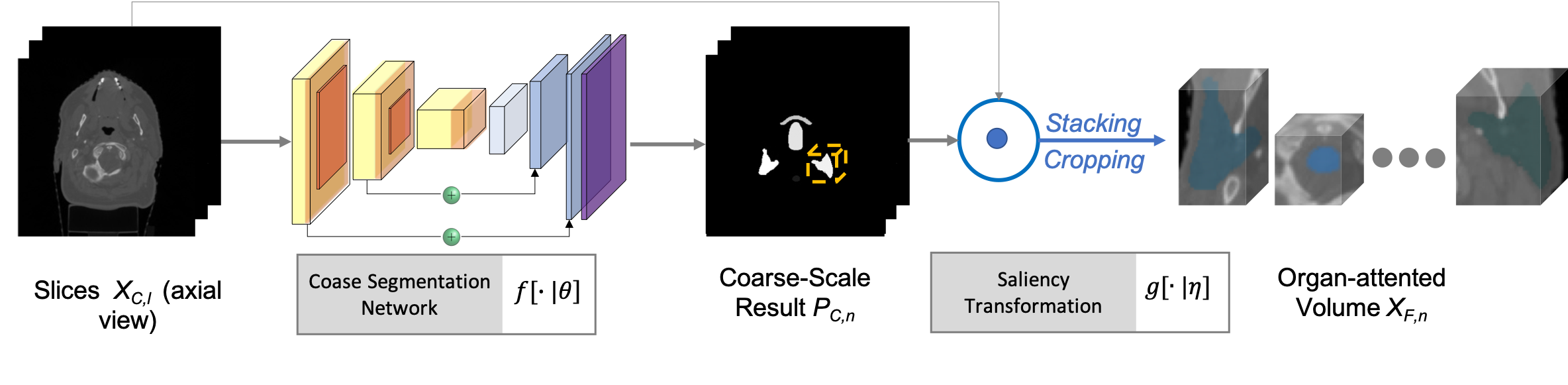}
    \caption{The sub-volume generator whose input CT volume $X_C$ is sliced along with axial view (z-axis)}
    \label{fig2-2}
\end{figure*}

The 3D sub-volume generator (shown in Fig. \ref{fig2-2}) involves a coarse segmentation network and a saliency transformation network \cite{JHU}, are fed with a sequence of 2D images by slicing input volume along with axial view and generate an organ-attended images slices. Cropping from the stack of the output images from the saliency transformation module based on prior bounding boxes, 3D organ-attended volumes, which are smaller and surrounded by simplified and less variable context, are generated as input to the 3D segmentation networks. 

We train a 2D deep network (2D-FCN \cite{FCN}) for coarse-scale segmentation, using Dice-Sorensen Coefficient (DSC) loss to prevent the model from being biased towards the background class. Each input 3D volume $X_C$ is sliced along the axial views, turning into a sequence of 2D slices denoted by $X_{C,l}(l = 1, 2, . . . , L)$ ($l$ denotes the index of slice and $L$ is the total number of slices along z-axis). The output of the network is a probability map $P_{C,l}$ (shown in Eq. \ref{eq2.2_1}, which can be converted into a segmentation mask after a filtering operation.
\begin{equation}
   P_{C,l} =f[ X_{C,l} | \theta],
    \label{eq2.2_1}
\end{equation}
where $f[\cdot | \theta]$ denotes the 2D-FCN segmentation network with network parameters $\theta$. To make use of the coarse-scale segmentation result as feature prior in the fine-scale stage, we introduce a Saliency Transformation module \cite{JHU}, which takes the probability map to generate an updated input image via $I_{C,l} = X_{C,l}\odot g(P_{C,l}; \eta)$. By this way, the transformation function adds spatial weights to the original input image. Thus, the segmentation process becomes:
\begin{equation}
    P_{C,l} =f(X_{C,l}\odot g(P_{C,l}|\delta)|\theta),
    \label{eq2.2_2}
\end{equation}
Here $g[\cdot; \delta]$ is the transformation function \cite{JHU} with parameters $\delta$, and $\odot$ denotes element-wise product. Finally, We concatenate all $P_{C,l}$ to reproduce a 3D organ-attended volume $P_C$ and obtain 3D sub-volume through cropping process according to prior bounding boxes (shown in Eq. \ref{eq2.2_3}).
\begin{equation}
    X_{F,n} =Crop[P_C|B_n],
    \label{eq2.2_3}
\end{equation}
where $X_{F,n}$ is the output sub-volume belonging to organ class $n-th$ and $Crop[\cdot]$ is the cropping function. 

\subsection{3D Segmentation Network}

Fed with volumes of diverse organs $X_{F,n} \in \bm{X_F} = \{X_{F,n}|n = 1,2,3,...,N\}$, the 3D segmentation network is supposed to recognise organ category and generate corresponding segmentation. Based on a 3D-Unet \cite{3du-net} architecture, a classification branch is introduced in our 3D segmentation network to prevent segmentation failure resulted from wrong organ recognition (shown in Fig. \ref{fig2-3}). Since sub-volume contains abundant local information, we introduce an global feature representation learned from prior bounding boxes (described in Section 3.1) to the 3D segmentation network. Therefore, the fine-scale segmentation process can be illustrated as:
\begin{equation}
    P_{F,n},Q_{F,n} =f(X_{F,n},B_n|\zeta),
    \label{eq2.3_1}
\end{equation}
where $P_{F,n}$ is the segmentation masks in fine-scale stage and $\zeta$ is the network parameters, $Q_{F,n}$ is the predict label for the organ. Finally, we filter the probability map by pre-defined threshold $h_n$ according to organ class index and generate the segmentation masks (shown in Eq. \ref{eq2.3_2}).
\begin{equation}
    Z_{F,n} =Sgn(P_{F,n}>h_n),
    \label{eq2.3_2}
\end{equation}
where $Sgn$ donates the Sign Function.

\section{Experiments and Results}

In this section, we test the proposed network on Public Domain Database for Computational Anatomy (\textbf{PDDCA}) version 1.4.1 and an institute dataset of H\&N CT images. 

\subsection{Datasets}
\noindent\textbf{Public Domain Database for Computational Anatomy (\textbf{PDDCA)}:}
The \textbf{PDDCA}, used in MICCAI 2015 Head and Neck Auto Segmentation Grad Challenge, contains 48 patient CT volumes with anisotropic pixel spacing ranging from 0.76 to 1.27 $mm$ and inter-slice thickness ranging from 1.25 to 3.0 $mm$. The resolution of each scanned sample is $512 \times 512 \times L$ ($L \in [98, 240]$ and $L$ is the number of slices along with the z-axis). 32 of the 48 samples in the \textbf{PDDCA} have integrated annotations of nine organ structures including brainstem, left and right parotid glands, left and right optic nerves, left and right submandibular glands, optic chiasm and mandible. For effectiveness of training and fair comparison, the remaining 16 samples which miss one or more structures are not used in our experiments. We randomly split the dataset into a training set containing 22 samples and a testing set including 10 samples.

\noindent\textbf{Instituted Head and Neck Dataset:}
Another dataset we used to train and evaluate our model is our instituted H\&N dataset. This database consists of 56 high-resolution CT scans of patient samples. The resolution of each scanned sample is $512 \times 512 \times L$ ($L \in [303, 392]$ and $L$ is the number of slices along with the z-axis. The slice thickness is $1.0 mm$. The dataset contains 18 organs, including left lens, right lens, spinal cord, left submandibular gland, right submandibular gland, larynx, thyroid, left parotid gland, right parotid gland, left globe, right globe, esophagus, brain stem, lips, oral cavity, and mandible. To the best of our knowledge, there is no published paper to investigate an automatic segmentation for all of these organs in their studies. The corpus for all samples is annotated by the radiation oncologist with over 20-year head-neck cancer treatment experience and approved for patient treatment.  We split the dataset into training data which contains 40 samples randomly selected among the whole dataset and testing data that includes the remained 16 samples. We repeat the dataset-splitting process four times to confirm the CT scan samples in both training and testing data vary each time for cross-validation purpose. 

\noindent\textbf{Data Augmentation}
To avoid overfitting problem and enhance robustness of the proposed model. We augmented both dataset by adding Gaussian noise with $\sigma = 0.5, 1, 1.5$ to original input and rescaling the input along with the corresponding annotations from $-10\%$, and $10\%$ to enlarge the dataset. Therefore, we used 192 and 240 variations to train the proposed model on \textbf{PDDCA} and the instituted dataset respectively.

\subsection{Evaluation Metrics}
To evaluate the proposed model, the Dice-Sorensen
coefficient (DSC), defined as Eq. \ref{eq-DSC}, is introduced to measure segmentation results for each body sample over two tasks.
\begin{equation}
	DSC(\mathcal{Y}, \mathcal{Z}) = \frac{2\times|\mathcal{Y}\cap\mathcal{Z}|}{|\mathcal{Y}| + |\mathcal{Z}|}. 
	\label{eq-DSC}
\end{equation}
In Eq. \ref{eq-DSC}, $\mathcal{Y}$ is the ground mask and $\mathcal{Z}$ is the prediction mask. This metric falls in the range of $[0, 1]$ with $1$ implying perfect segmentation.

\subsection{Experimental Results}
\begin{figure*}
    \centering
    \begin{minipage}[b]{1\linewidth}
    \includegraphics[width=\linewidth]{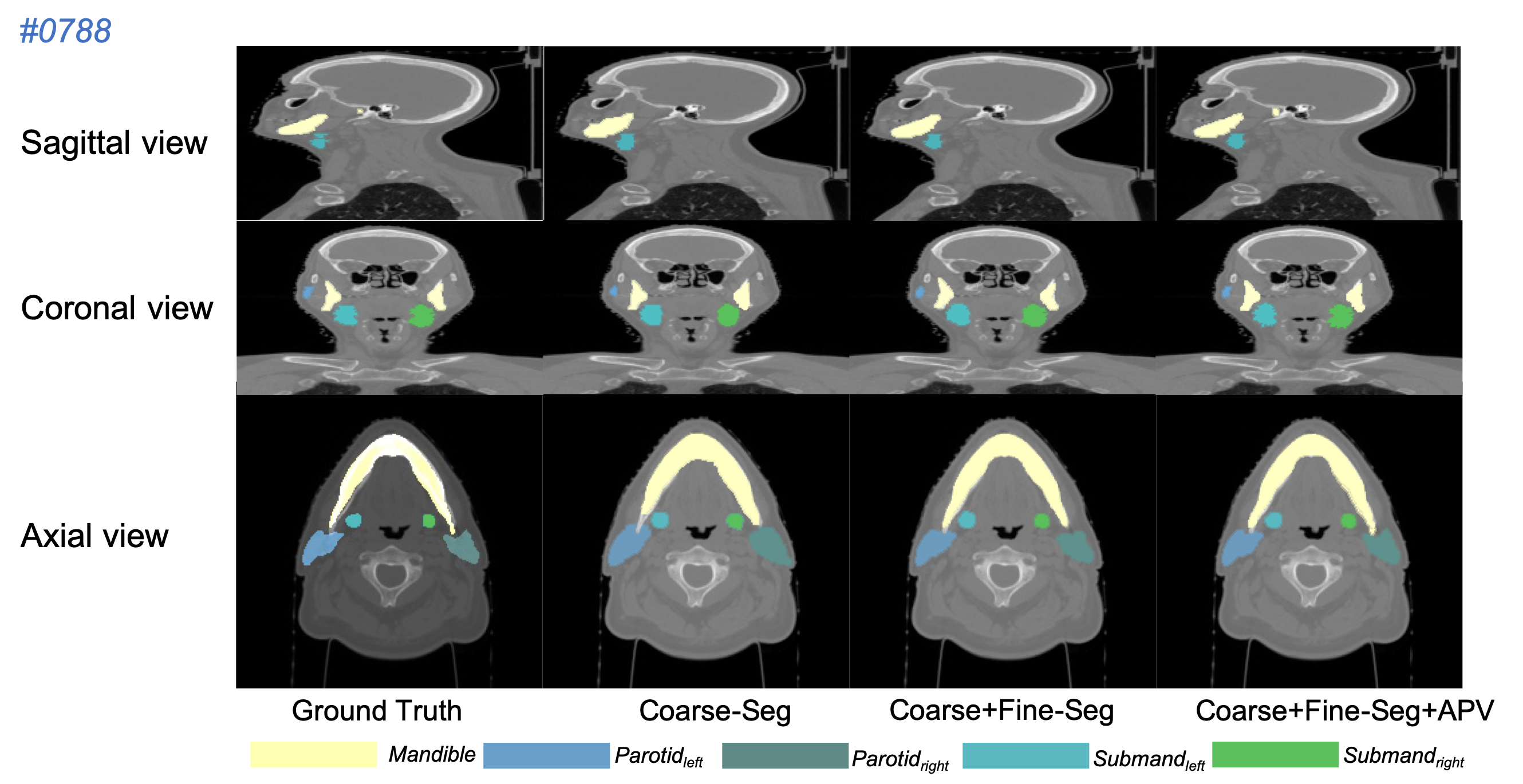}
    \end{minipage}
    \begin{minipage}[b]{1\linewidth}
    \includegraphics[width=\linewidth]{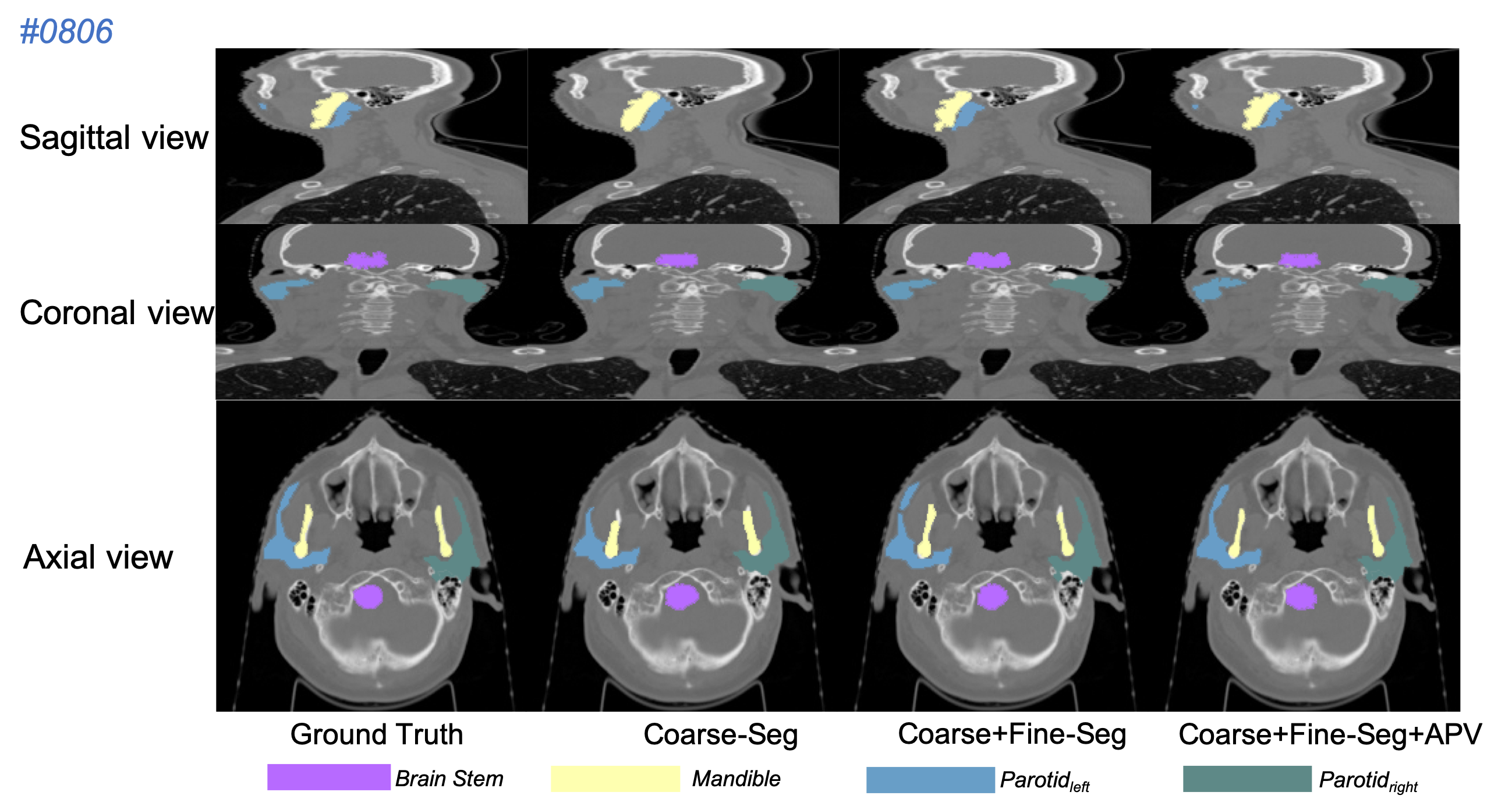}
    \end{minipage}
    \caption{Segmentation results of our proposed method on the \textbf{PDDCA} dataset.}
    \label{fig3-1}
\end{figure*}

The Quantitative comparison in Tab. \ref{tab4-1} demonstrates that the proposed network outperforms several state-of-the-art architectures based on the average DSC (\%) accuracy and is capable of segmenting organs from multi-classes robustly. Our coarse-to-Fine framework both without (\textbf{Our-F}) and with APV (\textbf{Our-A}) show competitive performance in direct comparison with the active appearance model method \cite{model_based} and current learning-based approaches \cite{SOAR,JHU}. Although \cite{JHU} shares similar coarse-to-fine architecture but processes only on 2D slices in fine-scale segmentation, our based-model (\textbf{Our-F}) directly exploits 3D volumes and learns structural information so as to achieve a higher accuracy. Enhanced by APV module, the refined segmentation (\textbf{Our-A}) gets highest DSC accuracy among all reported methods. Test results on patient samples randomly selected from the \textbf{PDDCA} validation dataset (Fig. \ref{fig3-1}) visually illustrate that the proposed method gives three-level segmentation: (1) the coarse-scale segmentation (Fig. \ref{fig3-1} Coarse-seg) predict a rough location and shape of the OARs; (2) the fine-scale segmentation (Fig. \ref{fig3-1} Coarse+Fine-seg) estimates a detail boundary of each OARs but still with some flaws; (3) the APV-refined segmentation (Fig. \ref{fig3-1} Coarse+Fine+APV-seg) generates 3D mask which is able to overlap the OARs perfectly.

\begin{table*}[h]
	\caption{Comparison between our Fine-scale segmentation (\textbf{Our-F}) and APV-refined segmentation (\textbf{Our-A}) and the previous approaches on the \textbf{PDDCA} dataset. Accuracy (DSC,$\%$) and standard deviation of all the OARs are reported. Higher accuracy and lower deviation are better.} 
	\begin{center}
		\begin{tabular}{p{3.5 cm}|p{2.3 cm}|p{2.1 cm}|p{2.1 cm}|p{2.2 cm}|p{2.2 cm}} \hline\hline
			\textbf{Organ}&Model-based \cite{model_based}&CNN\cite{SOAR}&CNN \cite{JHU}&Our-F&Our-A\\\hline
			Right Optic Nerves &63$\pm$5  &64.5$\pm$7.5  &62.4$\pm$6.7 &63.3$\pm$5.4 &\textbf{68.4$\pm$4.7}\\
			Left Optic Nerves  &63$\pm$5  &63.9$\pm$6.9  &62.1$\pm$9.2 &61.8$\pm$7.2 &\textbf{67.6$\pm$6.8}\\
			Right Submand      &78$\pm$8  &73.0$\pm$9.2  &74.2$\pm$6.0 &73.8$\pm$4.5 &\textbf{78.2$\pm$3.4}\\
			Left Submand       &78$\pm$8  &69.7$\pm$13.3 &68.9$\pm$8.2 &69.3$\pm$6.2 &\textbf{73.4$\pm$6.8}\\
			Optic Chiasm       &35$\pm$16 &37.4$\pm$13.4 &43.4$\pm$7.6 &47.5$\pm$7.3 &\textbf{54.7$\pm$8.2}\\
			Right Parotid      &82$\pm$10 &77.9$\pm$5.4  &84.3$\pm$3.1 &84.2$\pm$6.2 &\textbf{84.9$\pm$7.3}\\
			Left Parotid       &82$\pm$10 &76.6$\pm$6.1  &81.7$\pm$2.7 &82.5$\pm$4.1 &\textbf{84.2$\pm$5.0}\\
			Mandible           &91$\pm$2  &89.5$\pm$3.6  &86.4$\pm$2.9 &88.1$\pm$2.6 &\textbf{92.4$\pm$2.1}\\
			Brainstem          &\textbf{87$\pm$4}  &Unavailable   &81.2$\pm$2.3 &84.2$\pm$1.9 &84.9$\pm$1.9\\
			\hline\hline
		\end{tabular}
	\end{center}
	\label{tab4-1}
\end{table*}
 
\begin{table*}[h]
	\caption{Ablation study on the instituted Head-and-neck dataset. Accuracy (DSC,$\%$) of the coarse-scale (\textbf{Our-C}), the fine-scale (\textbf{Our-F}) and APV-refined (\textbf{Our-A}) segmentation are reported for each OARs. To better comprehend the performance, the final segmentation results of a current CNN-based method \cite{JHU} is also listed.} 
	\begin{center}
		\begin{tabular}{p{2.8cm}|p{2.2 cm}|p{2.2 cm}|p{2.2 cm}|p{2.7 cm}} \hline\hline
			\textbf{Organ}&\textbf{CNN \cite{JHU} }&\textbf{Our-C}&\textbf{Our-F}&\textbf{Our-A}\\\hline
			Right Submand  &54.48  &55.78& 56.96  & \textbf{57.53} \\
			Left Submand   &64.79  &63.42& 66.57 & \textbf{66.84} \\
			Right Parotid  &66.39  &66.20& 68.43 & \textbf{68.67} \\
			Left Parotid   &55.97  &51.37& 55.92 & \textbf{58.20}  \\
			Right Globe    &60.08  &60.91& 61.18 & \textbf{63.44} \\
			Left Globe     &61.23  &61.03& 61.66 & \textbf{63.58} \\
			Brainstem      &83.29  &85.04& 85.51 & \textbf{86.31}\\
			Lips           &\textbf{69.31}  &63.13& 66.10 & 67.63\\
			Oral Cavity    &83.10  &81.33& 82.25 & \textbf{82.91}\\
			Mandible       &87.73  &88.97& \textbf{91.78} & 90.48\\
			Larynx         &30.34  &29.87& 34.51 & \textbf{38.94}\\
			Esophagu       &68.74  &70.23& 72.66 & \textbf{73.47}\\
			Thyroid        &58.94  &58.20& 60.47 & \textbf{64.23}\\
			Spinal cord    &73.23  &75.42& 76.54 & \textbf{78.17}\\\hline
			\hline
		\end{tabular}
	\end{center}
	\label{tab4-2}
\end{table*}

\begin{figure*}
    \centering
    \includegraphics[scale=0.6]{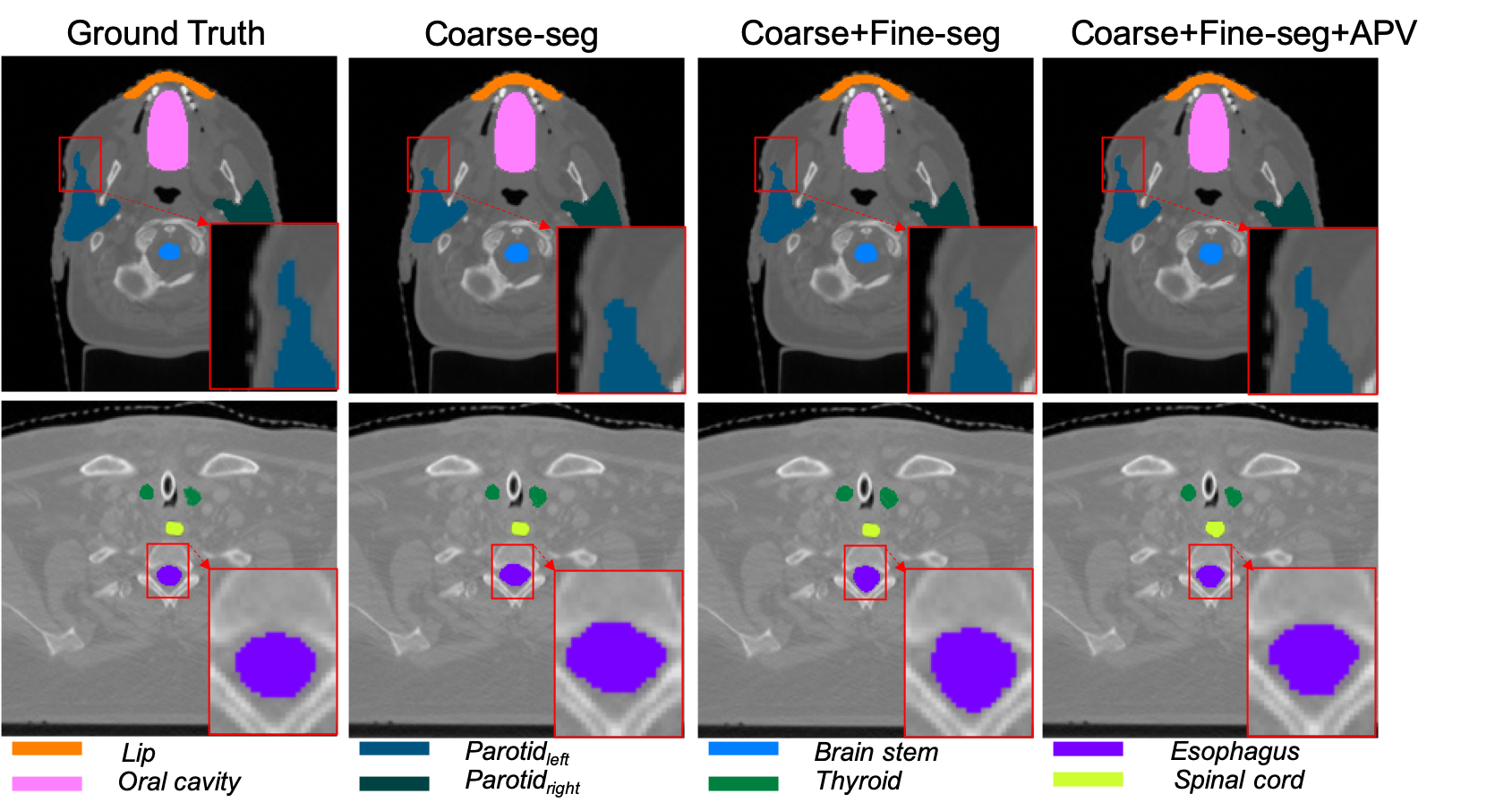}
    \caption{Example results of multi-organ segmentation on the instituted head-and-neck dataset.}
    \label{fig3-2}
\end{figure*}

Tab. \ref{tab4-2} illustrates an ablation study on the instituted H\&N dataset, where the accuracy (average DSC(\%)) of coarse-scale, fine-scale and APV refined segmentation are reported. Compared with \cite{JHU} that processes on 2D images in fine-scaled segmentation, our method has made a improvement ($1~2\%$ DSC) in the fine-scaled segmentation since it performs directly on 3D volume for all reported OARs, including some challenging organs (e.g. brain stem, esophagus, spinal cord and etc.) that are explored by little research. The relatively low accuracy in some OARs test is mainly caused by lack of organ annotations. For instance, only 6 over 56 body samples contain larynx annotation in ground truth so that all tests on this category shows a low accuracy. Moreover, it can be observed that our proposed method leads to a significant improvement ($1~5\%$) in segmentation precision for all considered organs by leveraging adversarial cues provided by the APV. Though all experimental results of different organs increase after applying the APV, the validator assists the segmentation network get a larger improvement (\textbf{Our-A}) when the base model (\textbf{Our-F}) has difficulty (i.e. $4\%$ increment in accuracy of Larynx segmentation). Meanwhile, as illustrated in Fig. \ref{fig3-2}, the qualitative measurement visually compares the segmentation results of the 8 H\&N OARs between the proposed base model without and with APV module. The refined segmentation by APV (\textbf{Our-A}) provides more integrated and consistent results, because the segmentation network is forced to segment details of OARs to ensure that no organ information is detected by APV module. The visual observation reflects that the validation network contributes to accurate segmentation in details.
\begin{table}[h]
	\caption{Segmentation Results on the Head-and-neck dataset. Accuracy (DSC,$\%$) comparison between our approach and the previous approach. Higher accuracy is better. 'Fine-seg' denotes fine scale segmentation, and 'Fine-seg(apv)' is fine scale segmentation with the validator.} 
	\begin{center}
		\begin{tabular}{l|c|c|c} \hline     \hline
			\textbf{Organ}&\textbf{3D UNet }&\textbf{UNet-based}&\textbf{JHU-based}\\\hline
			Right Submand &52.24& 54.23 & 57.53 \\
			Left Submand &67.24& 66.57 & 66.84 \\
			Right Parotid &67.34& 69.25 & 68.67 \\
			Left Parotid &59.24& 60.22 & 58.20  \\
			Right Globe &65.22& 65.79 & 63.44 \\
			Left Globe&59.10& 62.66 & 63.58 \\
			Brain stem &82.33& 85.61 & 86.31\\
			Lips &62.37& 65.75 & 67.63\\
			Oral Cavity&79.33& -& 82.91 \\
			Mandible &86.46&  - & 90.48\\
			Larynx &31.2& - & 38.94 \\
			Esophagus &71.29& - & 73.47 \\
			Thyroid &55.34& - & 64.23 \\
			Spinal cord &72.13& - & 78.17 \\\hline
			\hline
		\end{tabular}
	\end{center}
	\label{tab4-2}
\end{table}

\begin{table}[h]
	\caption{Segmentation Results on the NIH dataset. Accuracy (DSC,$\%$) comparison between our approach and the state-of-the-arts is reported. Higher accuracy is better. We only report the accuracy in Fine-scale segmentation. The subscript 'apv' represents accuracy of Fine-scale segmentation with the APV module.}
	\begin{center}
		\begin{tabular}{l|c|c|c} \hline 
		\hline
			\textbf{Approach}&\textbf{Mean}&\textbf{Max}&\textbf{Min}\\
			\hline
			Roth et al. \cite{NIH_com2} &  84.65$\pm$6.71\%     &86.29  & 23.99 \\
			Zhou et al. \cite{NIH_com3} &  82.37$\pm$5.58\%     &88.65  & 34.11\\
			Zhang et al. \cite{NIH_com5}&  77.89$\pm$8.52\%     &89.17  &43.67\\
			Roth et al. \cite{NIH_com4} &  81.27$\pm$6.27\%     &88.96  &50.69\\
			Zhou et al. \cite{NIH_com6} &  82.37$\pm$5.68\%     &90.85  &62.43\\
			Cai et  al. \cite{NIH_com1} &  82.40$\pm$6.70\%     &90.10  &60.00\\
			Yu   et al. \cite{JHU}      &  84.50$\pm$4.97\%     &91.02  &62.81\\
			Zhu   et al. \cite{JHU2}    &  84.59$\pm$4.86\%     &91.45  &69.62\\
			3D UNet                     &  82.64$\pm$1.24\%     &92.35  &59.97\\ \hline
			UNet-based                  &  83.98$\pm$2.17\%     &90.64  &64.11\\
			JHU-based                   &  87.50$\pm$4.89\%     &93.66  &62.19 \\\hline
			\hline
		\end{tabular}
	\end{center}
	\label{tab3-1}
\end{table}

\subsection{Implementation Details}

In general, we implemented our multi-organ segmentation with Adervisrial Performance Validator using the PyTorch \cite{pytorch} framework, which is an open-sourced deep learning platform that provides strong GPU support for computation efficiency.  We use SGD optimizer with a momentum of $0.9$ and an initial learning rate of $0.001$. We also use an early stopping mechanism to monitor training so that if the performance on test set does not improve for $10$ consecutive epochs then we decay the learning rate by half. The weighting parameters in total loss calculation are $\alpha = 2, \beta = 1$ and $\gamma = 0.5$, denoting weights of segmentation loss, classification loss and performance validation loss respectively. In the sub-volume generator module, one FCN-8s model \cite{FCN} pre-trained on PascalVOC \cite{Pascal} is applied as the base of the coarse-scale segmentation network. For the 3D segmentation network and the APV module, they share similar architecture based on a 3D U-net framework \cite{3du-net} and the same weights, except that APV involves no segmentation branch. The classification branches in 3D segmentation networks and APV involve dropout ratio of $0.5$ on the last fully connected layer before class prediction. We do not use dropout layer in any part of the segmentation branches in the sub-volume generator and 3D segmentation networks. Our model takes 15k iterations to converge on NIH Pancreas dataset with PyTorch using an NVIDIA RTX 2080 Ti GPU and the multi-organ segmentation takes around ten hours for 28k iterations to converge on the head-and-neck dataset with the same hardware setting. For fair comparison, we trained single-organ segmentation method \cite{JHU} for each OARs in the aforementioned datasets based on our prior bounding boxes.

\section{Discussion}

In this work, we proposed a 2D-3D hybrid segmentation network with a novel Adversaria Performance Validator (APV) for H\&N images to generate accurate 3D segmentation of OARs. The proposed network employs the APV module that refines segmentation by penalizing residual organ information after a soft masking process and compete with the segmentation network to further improve accuracy.

To verify the performance of the proposed network, comparison among a 2D coarse-to-fine network \cite{JHU}, our proposed 2D-3D hybrid base model and the based model with the APV is conducted and demonstrates: (1) fine-scale segmentation in 3D manner learns more sufficient geometric knowledge than the network \cite{JHU} that processes only 2D slices. By leveraging the prior location and shape information, the 3D network does not consume more computation resource compared with the 2D network with the same experimental setting. (2) The APV significantly improves the segmentation accuracy for organs with varying sizes, shapes, morphological complexities and CT contrasts in the way where segmentation network provides precise mask of an organ in volume renders to make it difficult for APV to recognize the organ. Although we observed that Performance Validation (PV) Loss will not decrease to zeros in our experiments because masking out a delineated organ preserve some shape information and neighboring anatomical context might imply the type of organ as well, the textural information underlying morphological complexities and CT contrasts of OARs provides significant indication of organ segmentation. Emphasising the segmentation loss by giving smaller weight of PV loss, the proposed network with APV achieves better performance than the base hybrid model.

Although the performance of compared methods can be influenced by the H\&N datasets. The hybrid network with the APV is shown to be competitive when compared with existing model-based and learning based methods by demonstrating consistently higher DSC (\%) and lower standard deviations. No post-processing involved when producing the competitive results, the proposed method has better generalizability.

\section{Conclusion}

Motivated by the limitation of 3D deep networks in CT scans organ segmentation due to large computational resource consumption. This work proposes a durable and effective strategy that jointly trains 2D segmentation and 3D segmentation networks in a coarse-to-fine manner. A novel Adversarial Performance Validator (APV) module is introduced into the pipeline to enhance the segmentation networks to generate refined organ-specific masks. Experimental results demonstrate segmentation networks with the APV is able to learn feature representation from local details more effectively.  In addition, we convert the information of the organ's location and shape over a large dataset into a global feature representation which assists segmentation network to better recognize a specific organ and learn organ feature representation more efficiently. The experimental results on the \textbf{PDDCA} dataset and our instituted head-and-neck dataset demonstrate the proposed network has a state-of-the-art performance on both single organ segmentation and multi-organ segmentation tasks.

{\small
\bibliographystyle{ieee_fullname}
\bibliography{MP}
}

\end{document}